\documentclass[nopreprintline,review,12pt]{elsarticle}

\usepackage{lineno,hyperref}
\modulolinenumbers[5]

\journal{Journal of Physics and Chemistry of Solids}

\usepackage{graphicx}
\usepackage{color}
\usepackage{booktabs}
\usepackage{tabularx, booktabs, makecell, caption}
\usepackage{siunitx}
\usepackage[T1]{fontenc}
\date{}

\usepackage{amssymb}
\usepackage{amsmath}

\usepackage{gensymb}
\usepackage{textcomp}
\usepackage[version=4]{mhchem}

\let\qty\SI

\begin{document}
\begin{frontmatter}

\title{Peculiarities of the local structure in new medium- and high-entropy, low-symmetry tungstates}

\author[ISSP]{Georgijs Bakradze\corref{mycorrespondingauthor}}
\cortext[mycorrespondingauthor]{Corresponding author}
\ead{georgijs.bakradze@cfi.lu.lv}

\author[DESY]{Edmund Welter}

\author[ISSP]{Alexei Kuzmin}

\address[ISSP]{Institute of Solid State Physics, University of Latvia,\\ Kengaraga 8, Riga, LV-1063, Latvia}

\address[DESY]{Deutsches Elektronen-Synchrotron DESY, Notkestrasse 85, 22607 Hamburg, Germany}

\begin{abstract}
\small
New monoclinic ($P2$/$c$) tungstates -- a medium-entropy tungstate, \ce{(Mn,Ni,Cu,Zn)WO4},  and a high-entropy tungstate, \ce{(Mn,Co,Ni,Cu,Zn)WO4}  -- were synthesized and characterized.
Their phase purity and solid solution nature were confirmed by powder X-ray diffraction and Raman spectroscopy.
X-ray absorption spectroscopy was used to probe the local structure around metal cations.
The atomic structures based on the ideal solid solution model were optimized by a simultaneous analysis of the extended X-ray absorption fine structure spectra at multiple metal absorption edges -- five for \ce{(Mn,Ni,Cu,Zn)WO4} and six for \ce{(Mn,Co,Ni,Cu,Zn)WO4} -- by means of reverse Monte Carlo simulations.
In both compounds, \ce{Ni^{2+}} ions have the strongest tendency to organize their local environment and form slightly distorted \ce{[NiO6]} octahedra, whereas \ce{Mn^{2+}}, \ce{Co^{2+}}, and \ce{Zn^{2+}} ions have a strongly distorted octahedral coordination.
The most intriguing result is that the shape of \ce{[CuO6]} octahedra in \ce{(Mn,Ni,Cu,Zn)WO4} and \ce{(Mn,Co,Ni,Cu,Zn)WO4} differs from that found in pure \ce{CuWO4}, where a strong Jahn--Teller distortion is present: \ce{[CuO6]} octahedra become more regular with increasing degree of dilution.
\end{abstract}


\end{frontmatter}


\newpage

\section{Introduction}\label{s:intro}
The concept of high-entropy materials (HEMs) -- first proven for metallic systems -- is now actively being transferred to other classes of non-metallic materials: carbides \cite{Hsieh2013118}, nitrides \cite{Braic2012117}, oxides \cite{Rost2015}, sulfides \cite{Zhang2018}, etc.
Although some HEMs are known to exhibit exceptional properties,  numerous fundamental questions are yet to be clarified \cite{SARKAR202043}.

The local chemical order is believed to play an important role in HEMs, although its influence on their macroscopic properties is not yet fully understood.
So far, several attempts have been made to investigate HEMs using techniques sensitive to the local atomic structure, such as X-ray absorption spectroscopy (XAS) \cite{Rost2015,Zhang2017,FANTIN2020,OH2021,Smekhova2022a,Smekhova2022b}.

For complex materials, XAS can provide unique information on the local atomic structure in terms of radial distribution functions (RDFs) \cite{Timoshenko2017a, DiCicco2018}.
The new opportunities in advanced XAS data analysis can be readily exploited to study monoclinic tungstates -- a promising class of functional materials \cite{Timoshenko2014,Bakradze2021}. Recently, a high-entropy molybdate -- a representative of a related class of materials -- was investigated \cite{Stenzel2021} but no attempts to synthesize and investigate high-entropy tungstates (HETs) have been published up to now.

Versatile physical and chemical properties of tungstates find a wide range of applications in scintillators \cite{Millers1997, Nikl2008}, down-conversion phosphors \cite{Vanetsev2016}, white light-emitting diodes \cite{Zhai2016}, supercapacitors \cite{Niu2013}, lithium-ion batteries \cite{Wang2018}, and laser host materials \cite{Pask2003}. Recently their use as heterogeneous catalysts \cite{Zhang2014, Yan2019}, humidity or gas sensors \cite{Bhattacharya1997}, electrochromic materials \cite{Kuzmin2001}, anticorrosion pigments \cite{Kalendova2015}, and optical temperature sensors \cite{Zhang2018} and in optical recording \cite{Kuzmin2007, Kuzmin2015} has been explored too. Many functional properties of tungstates can be further modified by reducing crystallite size, doping, or making solid solutions with other tungstates \cite{Bakradze2021,Dey2014,Kuzmin2016}. The solid solution approach is of particular interest because of a wide range of possible chemical compositions.

\begin{figure}[t]
	\centering	
	\includegraphics[width=0.75\textwidth]{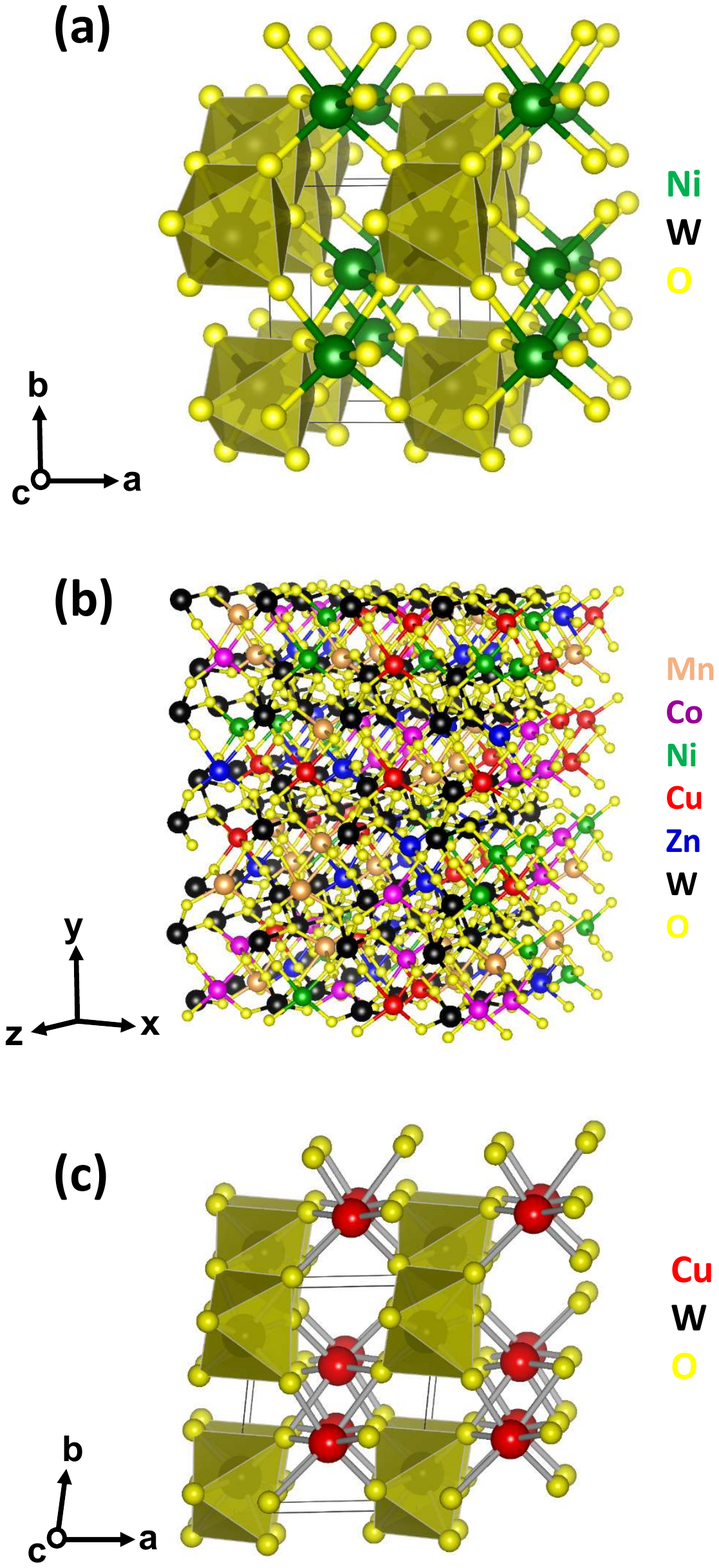}	
	\caption{Crystallographic structures of (a) monoclinic \ce{NiWO4}, (b) monoclinic 	\ce{(Mn,Co,Ni,Cu,Zn)WO4}, and (c) triclinic \ce{CuWO4}. See the main text for details.}
	\label{fig1}
\end{figure}

\ce{$A$WO4} tungstates -- where \ce{$A$} is a small divalent cation such as \ce{Mn}, \ce{Co}, \ce{Ni}, or \ce{Zn} -- crystallize in the wolframite crystal structure, space group $P2$/$c$, no. 13; see Fig.\ \ref{fig1}(a) \cite{Sleight1972}.
The wolframite structure can be thought of as a hexagonal close-packed array of oxygen anions with the tungsten and metal cations in the octahedral holes, forming infinite independent zigzag chains along the [001] direction in the crystal; in the perpendicular [100] direction, the short chains of \ce{[$A$O6]} octahedra form a layer that alternates with a layer of edge-joined \ce{[WO6]} octahedra.
Each chain of \ce{[WO6]} octahedra is attached by common corners to four chains of \ce{[$A$O6]} octahedra, and vice versa.
The \ce{[$A$O6]} octahedra are slightly distorted, whereas the \ce{[WO6]} octahedra are strongly distorted, with tungsten ions being located off-centre because of the second-order Jahn--Teller (JT) effect induced by the \ce{W^{6+}}~(5d$^0$) electronic configuration \cite{KUNZ1995}.

In triclinic \ce{CuWO4} (Fig.\ \ref{fig1}(c)), the electronic configuration of \ce{Cu^{2+}}~(3d$^9$) cations results in a strong first-order JT effect, leading to an axial distortion of \ce{[CuO6]} octahedra \cite{Forsyth1991}.
Such distortion lowers the crystal symmetry from monoclinic to triclinic ($P\bar{1}$, no. 2).
The lattice distortion induced by the JT effect leads to rather small variations of the lattice parameters and positions of atoms in the unit cell \cite{Sleight1972}.
As a result, tungstates with monoclinic and triclinic lattices can easily mix and form solid solutions \cite{Simon1996,Naik2009,Selvan2009,Yourey2012}.
Nevertheless, differences in the sizes and electronic configurations of cations contribute to a change in their local atomic structure, which leads to various distortions of their coordination octahedra.
Therefore, it is interesting to investigate how local environments of \ce{$A$^{2+}} cations will respond to the formation of a high-entropy \ce{(Mn,Co,Ni,Cu,Zn)WO4} complex  with compositional disorder.

In this study,  we synthesized a new medium-entropy tungstate (MET), \ce{(Mn,Ni,Cu,Zn)WO4}, and a new HET, \ce{(Mn,Co,Ni,Cu,Zn)WO4}, and investigated their structure using powder X-ray diffraction (XRD), Raman spectroscopy, and multi-edge XAS combined with reverse Monte Carlo (RMC) simulations.
Our results confirm the formation of a single-phase MET and a single-phase HET, and shed light on local structure distortions around metal cations.

\section{Experimental and data analysis}\label{s:exper}
\ce{(Mn,Ni,Cu,Zn)WO4} and \ce{(Mn,Co,Ni,Cu,Zn)WO4} were synthesized by a co-precipitation method from stoichiometric amounts of \ce{Na2WO4.2H2O} and 3d metal nitrates as educts. First, each salt was dissolved in \qty{20}{\milli\litre} of deionized water.
Next, the solutions containing 3d metal salts were mixed at room temperature and slowly added to the tungstate aqueous solution at pH 8. The light brown precipitates were collected, thoroughly washed with water, acetone, and 2-propanol, and annealed at \qty{550}{\celsius} for \qty{5}{\hour} in air, resulting in ochre-coloured powders.

To determine the chemical composition of the  samples obtained, inductively coupled plasma (ICP) mass spectrometry measurements were conducted (Thermo Fischer model iCAP TQe). A multi-element standard was used for external calibration (ICP multi-element standard solution IV, Merck, Germany).
To prepare analyte solutions, about \qty{50}{\milli\gram} of the samples was dissolved in \qty{30}{\milli\gram} of hydrochloric acid and \qty{10}{\milli\gram} of nitric acid at \qty{400}{\kelvin} for 30~min. The chemical analysis was accomplished with four different calibration solutions and an internal standard. The oxygen content was determined by the missing weight. The analysis yielded the following atomic concentrations: Mn 3.7~at.\%, Ni 4.2~at.\%, Cu 3.9~at.\%, Zn 4.0~at.\%, W 16.8~at.\%, and O 67.4~at.\% for the MET, and Mn 2.7~at.\%, Co 3.1~at.\%, Ni 3.2~at.\%, Cu 3.2~at.\%, Zn 3.0~at.\%, W 17.0~at.\%, and O 67.8~at.\% for the HET, thus confirming the nominal stoichiometric compositions of the samples.

\begin{figure}[t]
	\centering	
	\includegraphics[width=0.5\textwidth]{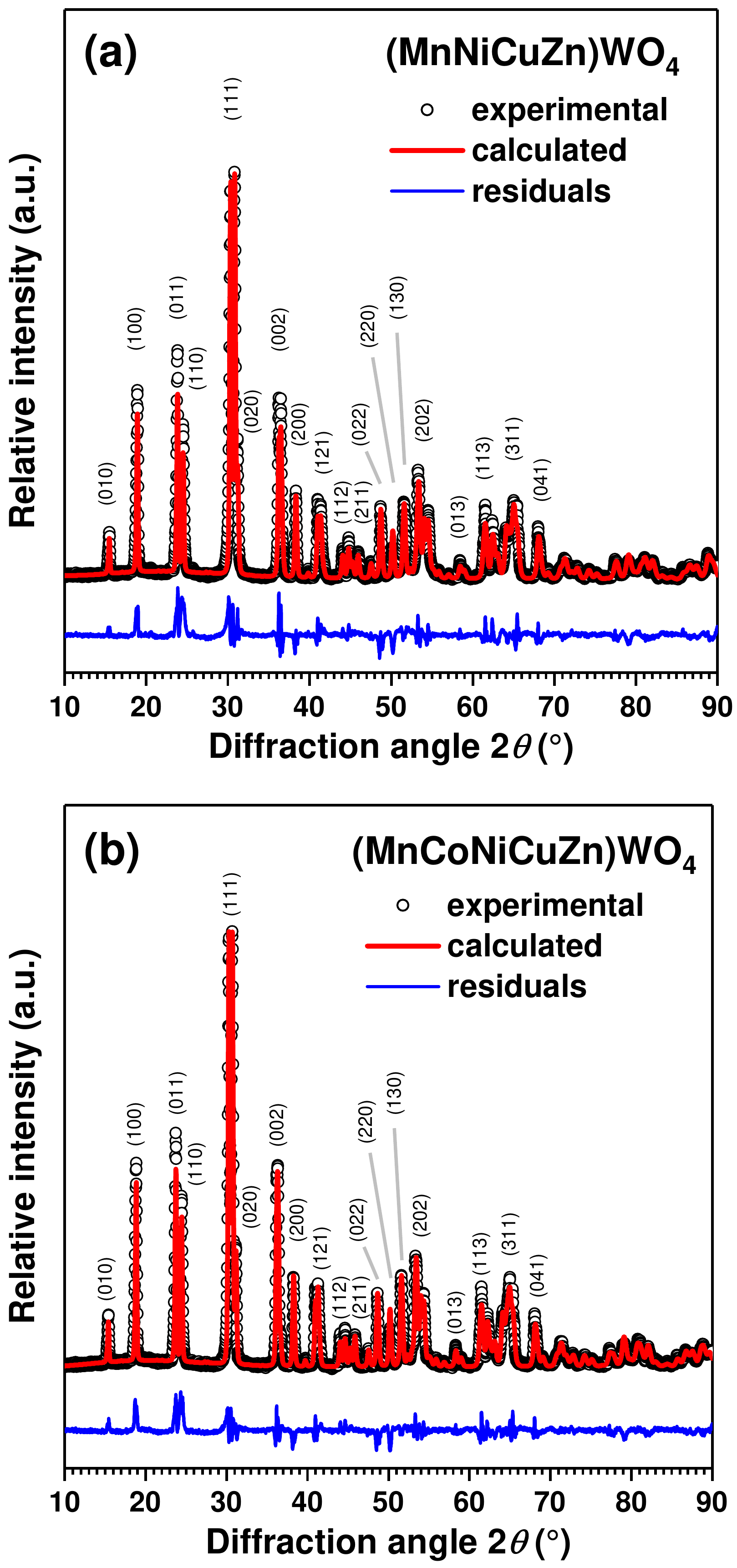}	
	\caption{Experimental and calculated X-ray diffraction patterns of (a) medium-entropy tungstate  \ce{(Mn,Ni,Cu,Zn)WO4} and (b) high-entropy  tungstate \ce{(Mn,Co,Ni,Cu,Zn)WO4}. See the main text for details.}
	\label{fig2}
\end{figure}

The phase purity of all samples was checked by powder XRD at room temperature with use of a benchtop diffractometer with a Bragg--Brentano geometry (Rigaku MiniFlex 600, Cu K$_\alpha$ radiation, operated at \qty{40}{\kilo\volt} and \qty{15}{\milli\ampere}).
Fig.\ \ref{fig2} shows the experimental XRD patterns for the MET and the HET. The qualitative analysis  and quantitative refinement of the diffraction data were performed with Profex version 4.3.6 \cite{Doebelin2015}.

\begin{figure}[t]
	\centering
	\includegraphics[width=1\textwidth]{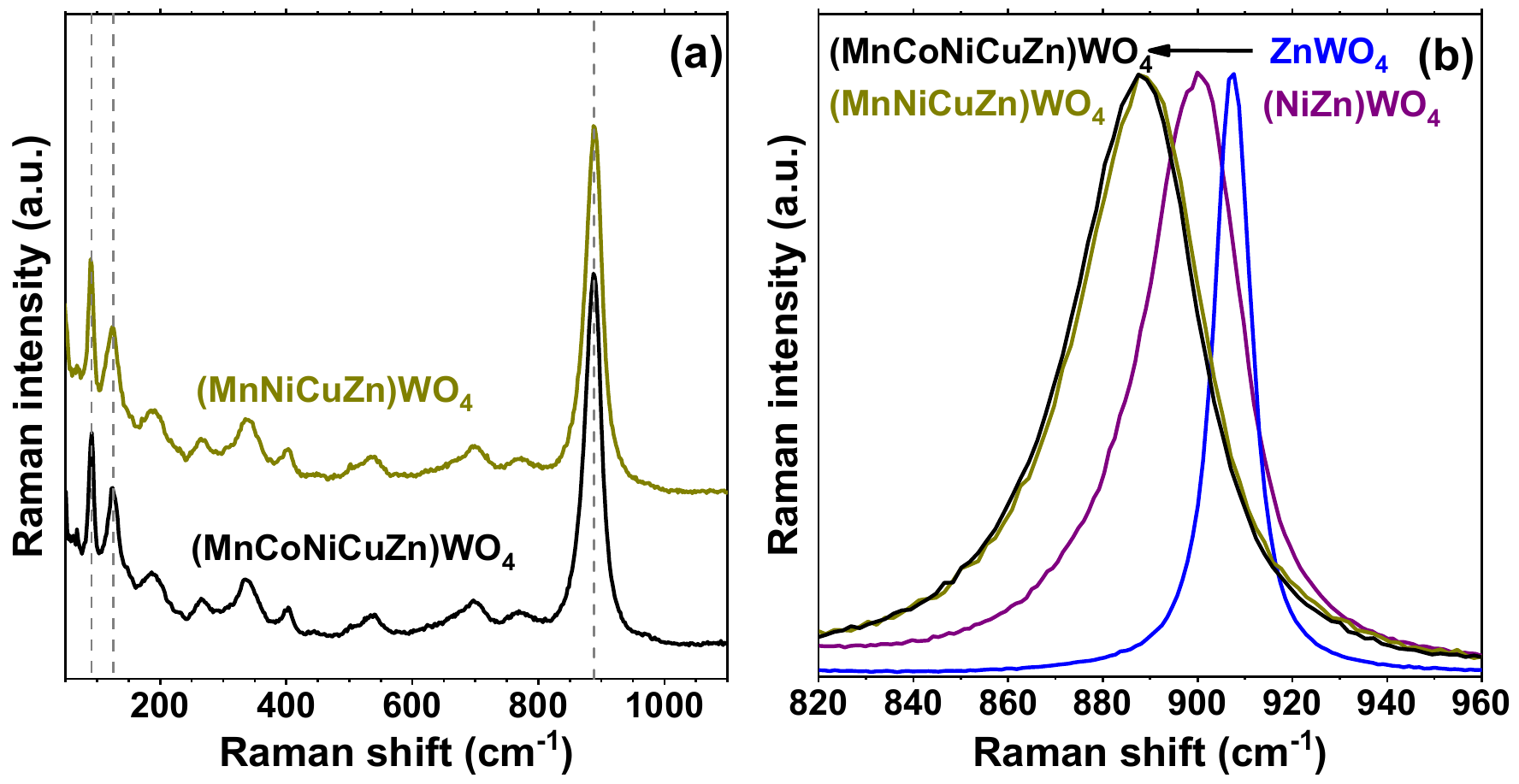}
	\caption{Raman scattering spectra of medium-entropy tungstate \ce{(Mn,Ni,Cu,Zn)WO4} and high-entropy tungstate \ce{(Mn,Co,Ni,Cu,Zn)WO4}: (a) full range and (b) $\mathrm{A_g}$ band region. Spectra of \ce{ZnWO4} and \ce{(Ni,Zn)WO4} are shown for comparison in (b).}
	\label{fig3}
\end{figure}

Micro-Raman spectroscopy measurements were performed in a backscattering geometry with a TriVista 777 confocal Raman microscopy system (Princeton Instruments, \qty{750}{\milli\metre} focal length, grating with 600~lines per millimetre).
A Cobolt Samba 150 \qty{532}{\nano\metre} continuous-wave single-frequency diode-pumped solid-state laser  was used for excitation through an Olympus UIS2 UPlanFL N $20\times$/0.50 objective.
The recorded Raman spectra are shown in Fig.\  \ref{fig3}.
The data for \ce{ZnWO4} and \ce{(Ni,Zn)WO4} were taken from our previous work \cite{Bakradze2020, Bakradze2021}, and are shown for comparison.

XAS measurements were conducted at the  HASYLAB PETRA III P65 undulator beamline \cite{Welter2019}. The PETRA III storage ring operated at $E= \qty{6.08}{\giga\electronvolt}$ and $I=\qty{120}{\milli\ampere}$ in a top-up 480 bunch mode.
The harmonic rejection was achieved by use of uncoated silicon plane mirrors. Fixed exit Si(111) and Si(311) monochromators were used.
The X-ray absorption spectra at the K-edge of Mn (6539 eV), Co (7709 eV), Ni (8333 eV), Cu (8979 eV), and Zn (9659 eV) and the L$_3$-edge of W (10200 eV) were collected in transmission mode at \qty{10}{\kelvin} with use of a liquid helium cryostat.
Microcrystalline powder samples were deposited onto Millipore filters and fixed with Scotch tape.
The sample weight was chosen to result in absorption edge jumps of around 1.

The extended X-ray absorption fine structure (EXAFS) spectra $\chi(k)k^2$ of the MET (HET) at five (six) absorption edges were extracted from the measured XAS spectra with use of a conventional procedure \cite{Kuzmin2014}.
The complex structure and the low symmetry of tungstates complicate significantly the interpretation of EXAFS spectra by conventional methods, especially for the outer coordination shells \cite{Timoshenko2014,Bakradze2021}; therefore, we analysed the EXAFS spectra using the RMC method and an evolutionary algorithm approach as implemented in the EvAX code \cite{Timoshenko2014rmc}.
With this approach, a model of the atomic structure is built, the shape and size of which remain unchanged during the simulation process.
After each iteration, atomic positions are varied randomly to account for static and dynamic disorder in the material. Theoretically calculated structure-related data (i.e. in our case EXAFS spectra at each edge: five in the case of the MET and six in the case of the HET) are recalculated and compared with experimental spectra.
Only those changes in atomic positions that result in a better fit between experimental and theoretically calculated EXAFS spectra are accepted.
Each RMC simulation results in one structural model; however, simulations are repeated many times starting from different initial atomic offsets to accumulate the statistics.

\begin{figure*}[t]
	\centering
	\includegraphics[width=1\textwidth]{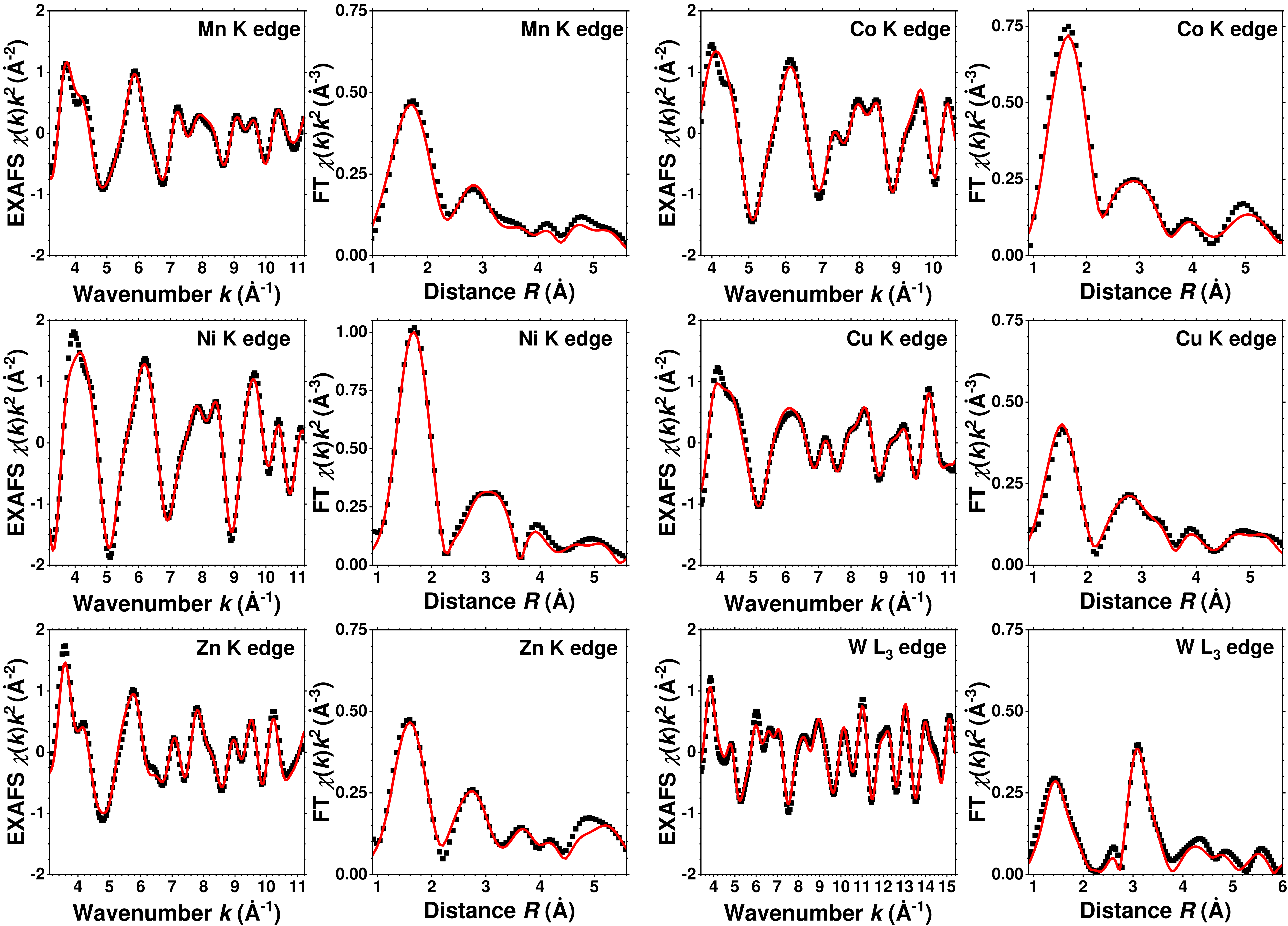}	
	\caption{Experimental (black dots) and calculated (red lines) extended X-ray absorption fine structure (EXAFS) spectra $\chi(k)k^2$ and their Fourier transforms (FTs) for high-entropy tungstate \ce{(Mn,Co,Ni,Cu,Zn)WO4} at the Mn, Co, Ni, Cu, and Zn K-edges and the W L$_3$-edge at \qty{10}{\kelvin}.
	The peak positions in the FTs differ from their true crystallographic values because of the EXAFS phase shifts. See the main text for details.}
	\label{fig4}
\end{figure*}

Initial structural models for the HET were constructed as $4a \times 4b \times 4c$ supercells -- where $a$, $b$, and $c$ are the lattice parameters determined from the XRD data -- with a random distribution of 3d cations at the $A$ sites in the required proportions.
In our model, all 3d cations are homogeneously distributed since tungstates are known to easily form solid solutions \cite{Simon1996,Naik2009,Selvan2009,Yourey2012}, and we do not expect to have any appreciable segregation of metal cations in our samples at the nanoscale.
In the fitting procedure, all atoms within a supercell were randomly displaced at each iteration, with the maximum allowed displacement being \qty{0.4}{\angstrom}.
Theoretical EXAFS spectra at five metal edges for the MET or six metal edges for the HET  were calculated with use of the ab initio self-consistent real-space multiple scattering FEFF8.5L code \cite{Ankudinov1998} and compared with experimental spectra in the direct ($R$) and reciprocal ($k$) spaces simultaneously.
Good agreement was achieved after several thousand RMC iterations for both the MET and the HET; the fitting results for the HET are shown in Fig.\ \ref{fig4} as an example.

Atomic configurations obtained from RMC simulations were used to calculate partial RDFs for metal--oxygen atomic pairs, corresponding to the first coordination shell of metal atoms.
To improve statistics, RMC simulations were performed 72 times starting from eight different initial atomic configurations and using nine different series of pseudo-random numbers (i.e. the RDFs shown in Fig.\ \ref{fig5} are averaged over 72 different RDFs; the data given in Table\ \ref{table1} represent the average values over all unique octahedra in 72 final atomic structures).

In our approach, EXAFS data from multiple absorption edges (e.g. from the W L$_3$-edge and K-edges of 3d metals) are used simultaneously to optimize a single, self-consistent structural model to obtain an unambiguous solution.

\section{Results and discussion}\label{s:results}
The XRD patterns of \ce{(Mn,Ni,Cu,Zn)WO4} and \ce{(Mn,Co,Ni,Cu,Zn)WO4} were indexed and refined with use of a structural model of a monoclinic crystal (space group $P2$/$c$, no. 13) assuming equal occupancy probabilities for 3d metal cations at $A$ sites (Fig.\ \ref{fig2}). A good match between the experimental and refined XRD patterns evidences a single-phase monoclinic material with the following lattice parameters:
$a = (4.7018 \pm 0.0002)$~\AA,
$b = (5.7392 \pm 0.0002)$~\AA,
$c = (4.9286 \pm 0.0002)$~\AA, and
$\beta = (91.006 \pm 0.003)\degree$ for the MET and
$a = (4.6933 \pm 0.0002)$~\AA,
$b = (5.7272 \pm 0.0003)$~\AA,
$c = (4.9311 \pm 0.0002)$~\AA, and
$\beta = (91.227 \pm 0.003)\degree$ for the HET.

\begin{figure}[t]
	\centering
	\includegraphics[width=0.5\textwidth]{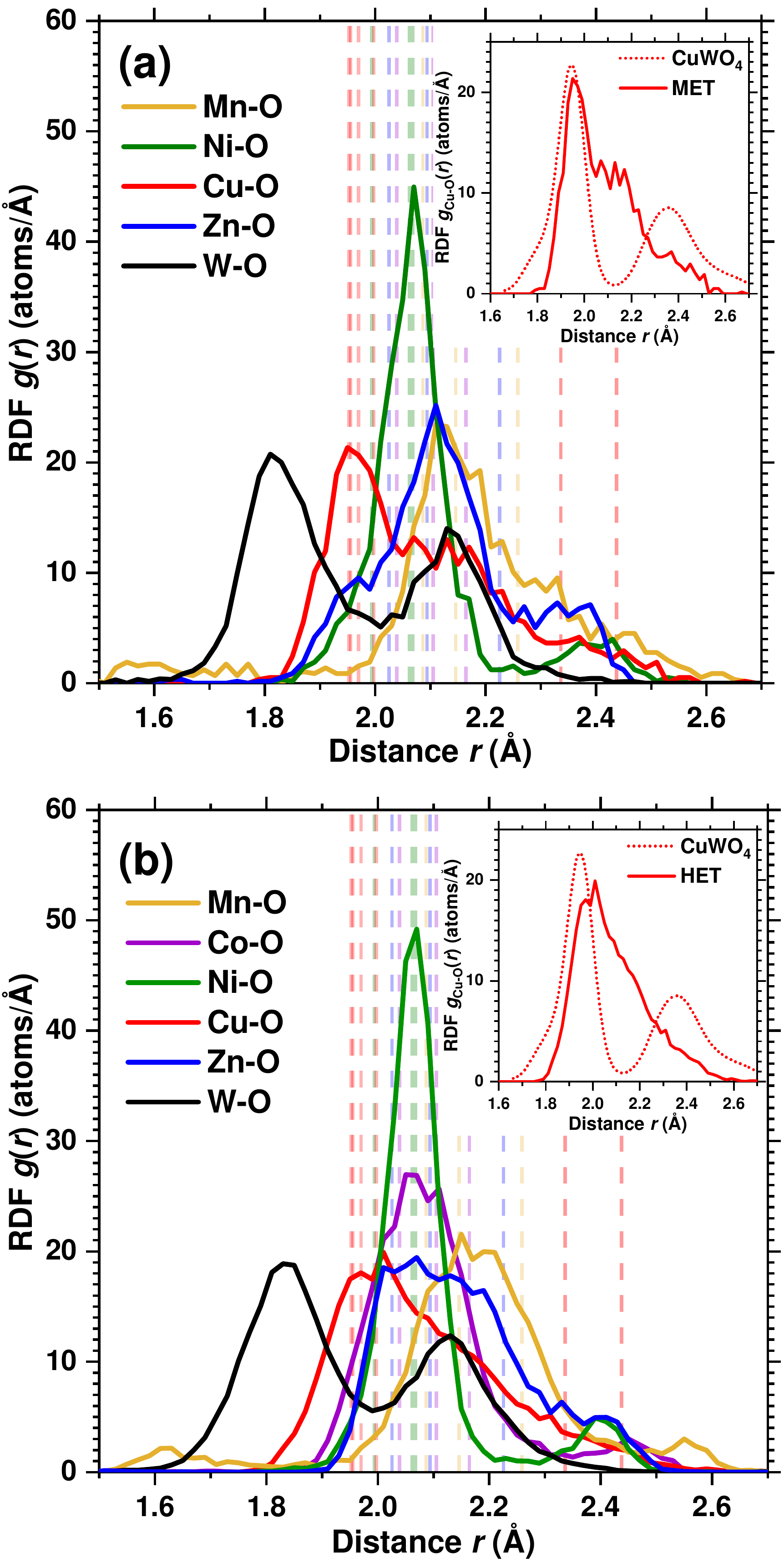}	
	\caption{Radial distribution functions (RDFs) $g(r)$ in (a) medium-entropy tungstate (MET) \ce{(Mn,Ni,Cu,Zn)WO4} and (b) high-entropy tungstate (HET) \ce{(Mn,Co,Ni,Cu,Zn)WO4}  at \qty{10}{\kelvin}. Vertical dashed lines indicate the respective bond lengths in pure \ce{MnWO4}, \ce{CoWO4}, \ce{NiWO4}, \ce{CuWO4}, and \ce{ZnWO4} as determined by X-ray diffraction at \qty{300}{\kelvin}. The inset shows partial RDFs for Cu--O atomic pairs in \ce{CuWO4} and \ce{(Mn,Ni,Cu,Zn)WO4} or \ce{(Mn,Co,Ni,Cu,Zn)WO4}. See the main text for details.}
	\label{fig5}
\end{figure}

The Raman spectra of \ce{(Mn,Ni,Cu,Zn)WO4} and \ce{(Mn,Co,Ni,Cu,Zn)WO4} are qualitatively  similar, and consist of a set of peaks due to the Raman-active even $\mathrm{A_g}$ modes (Fig.\ \ref{fig3}(a)).
According to group theory \cite{Kuzmin2011niwo4,Kuzmin2013}, there are 18 Raman-active modes  in the wolframite structure.
A clear shift of the most intense $\mathrm{A_g}$ band positioned at around \qty{900}{\per\cm} \cite{Bakradze2020,Wang1992} is apparent in Fig.\ \ref{fig3}(b) (the Raman spectra of \ce{ZnWO4} and \ce{(Ni,Zn)WO4} are also shown for comparison).
Note that the tungstates shown in Fig.\ \ref{fig3}(b) were chosen according to the gradual increase in the number of different 3d cations in their chemical formula.
The $\mathrm{A_g}$ band corresponds to the symmetric stretching W--O mode of a \ce{[WO6]} octahedron \cite{Errandonea2018}. We showed earlier that the position and width of the $\mathrm{A_g}$ band are sensitive to distortions of \ce{[WO6]} octahedra and the W--O bond strength \cite{Bakradze2021,Bakradze2020}.
The frequency of the $\mathrm{A_g}$ band decreases (Fig.\ \ref{fig3}(a)) by about \qty{8}{\per\cm} in the series  \ce{ZnWO4}, \ce{(Ni,Zn)WO4}, \ce{(Mn,Ni,Cu,Zn)WO4}, and \ce{(Mn,Co,Ni,Cu,Zn)WO4} (i.e. with increasing configurational entropy), and reaches its minimum value in the HET at around \qty{889}{\per\cm}. Thus, the related W--O bonds in the HET are the weakest.
The changes in vibrational dynamics can be explained by competing $A$--O and W--O interactions resulting in a variation of the W--O bond strength \cite{Bakradze2021,Bakradze2020}: Zn--O bonds are more ionic than other $A$--O bonds because of the closed electronic configuration of \ce{Zn^{2+}} ions; thus, the gradual replacement of zinc with other 3d elements reduces the degree of covalency of W--O bonds, making them weaker. The broad shape of the $\mathrm{A_g}$ band indicates multiple atomic configurations existing around W atoms in the HET (see the RDFs for the  W--O atomic pair shown in Fig.\ \ref{fig5}).

The advanced analysis of the low-temperature EXAFS spectra at five metal absorption edges for \ce{(Mn,Ni,Cu,Zn)WO4} and six metal absorption edges for \ce{(Mn,Co,Ni,Cu,Zn)WO4} simultaneously using the RMC method allowed us to optimize the atomic configurations  on the basis of the ideal solid solution model and to study local static distortions around each metal cation in both compounds.
An example of the  structure obtained for \ce{(Mn,Co,Ni,Cu,Zn)WO4} is shown in Fig.\ \ref{fig1}(b).

The RDFs for metal--oxygen atomic pairs are shown in Fig.\ \ref{fig5}. The W--O RDF exhibits two peaks located at \qty{1.83}{\angstrom} and \qty{2.13}{\angstrom}, just as in pure wolframites \cite{Timoshenko2014,Bakradze2021}, where the distortion is caused by the second-order JT effect for W$^{6+}$~(5d$^0$) \cite{KUNZ1995}.
At the same time, the formation of the HET has a pronounced effect on \ce{[$A$O6]} structural units: the Mn--O and Co--O RDFs are broad and symmetric, whereas the Zn--O RDF exhibits a strongly asymmetric shape with a tail on the long-distance side.
The average Mn--O distance is considerably greater than the other $A$--O distances because a sixfold-coordinated Mn$^{2+}$ in the high-spin state has the largest ionic radius among the cations considered (\qty{0.83}{\angstrom} \cite{Shannon1976}).

In contrast, the Ni--O RDFs in both compounds are narrow and symmetric.
This fact stresses the structural role of nickel ions, which have the ability to organize their local environment into regular \ce{[NiO6]} octahedra, while other $A$-type ions simply adapt to it (cf. the behaviour of Ni in \ce{Zn_cNi_{1-c}WO4} solid solutions \cite{Bakradze2021}).

The most intriguing result is the shape of the Cu--O RDF (see Fig.\ \ref{fig5}): in the MET, it is strongly asymmetric, with a maximum at around \qty{2.00}{\angstrom} and a distinct shoulder at around \qty{2.15}{\angstrom}; in the HET, it is also strongly asymmetric, with a maximum at around \qty{2.00}{\angstrom} and a long tail on the long-distance side.
In both cases, the shape of the Cu--O RDF curves significantly deviates from that in pure \ce{CuWO4} (see the inset in Fig.\ \ref{fig5}(b) or Fig.~5 in Ref. \cite{Timoshenko2014} for details), where the RDF splits into two distinct peaks corresponding to the nearest four in-plane oxygen atoms at around \qty{2.00}{\angstrom} and the remaining two axial oxygen atoms at around \qty{2.30}{\angstrom}.
Therefore, one can conclude that the increase in the configurational entropy indirectly (i.e., by creation of a local environment with many chemically different neighbours) significantly affects the longest Cu--O bonds in \ce{[CuO6]} octahedra.

The degree of distortion  of  \ce{[$A$O6]} octahedra can be quantified by the so-called JT distortion parameter, defined as
\begin{equation}
	\sigma_\text{JT} = \sqrt{\frac{1}{6} \sum_{i=1}^6 \left(r_i - \langle r_i \rangle \right)^2},
\end{equation}
where $r_i$ are the six $A$--O distances within an \ce{[$A$O6]} octahedron and $\langle r_i \rangle$ is the average $A$--O distance.
The JT distortion parameter values for different octahedra in the MET and the HET are given in Table~\ref{table1}.

\begin{table}[t]
	\caption{Jahn--Teller distortion parameter, $\sigma_\text{JT}$ (\AA), for different octahedra in medium-entropy tungstate (MET) \ce{(Mn,Ni,Cu,Zn)WO4} and high-entropy tungstate (HET) \ce{(Mn,Co,Ni,Cu,Zn)WO4} at 10 K.}
	\label{table1}
	
	\begin{tabular}{lll}
		\toprule
						& MET				& HET     \\
		\midrule
		\ce{[MnO6]} 	& $0.163 \pm 0.092$ & $0.152 \pm 0.090$\\
		\ce{[CoO6]} 	& --    				& $0.102 \pm 0.052$\\
		\ce{[NiO6]} 	& $0.091 \pm 0.057$	& $0.085 \pm 0.056$\\
		\ce{[CuO6]} 	& $0.140 \pm 0.051$	& $0.131 \pm 0.049$\\
		\ce{[ZnO6]} 	& $0.122 \pm 0.046$	& $0.115 \pm 0.043$\\
		\ce{[WO6]}  	& $0.159 \pm 0.033$	& $0.161 \pm 0.038$\\
		\bottomrule
	\end{tabular}
\end{table}

Thus, with increasing degree of dilution (i.e., with decreasing concentration) of 3d metal ions in the MET versus the HET, the JT distortion parameter decreases by about 6\%.
The degree of distortion of \ce{[WO6]} octahedra stays almost the same.
Our results indicate that the dilution has an effect qualitatively similar to that of applying hydrostatic pressure to pure tungstates: in both cases, the JT distortion parameter decreases (i.e. octahedra become less distorted).
In particular, in pure \ce{CuWO4} -- because of the different linking of octahedral units along different crystallographic directions -- \ce{[CuO6]} octahedra were found to be much more compressible than \ce{[WO6]} octahedra \cite{RuizFuertes2010}.
Similar effects have been observed in \ce{MnWO4} and \ce{ZnWO4} \cite{MacaveiSchulz1993, RuizFuertes2010a}.

It has been shown that in binary tungstates the $A$-type solute atoms often substitute directly for $A^\prime$-type atoms in the \ce{[$A^\prime$O6]} chains \cite{Bakradze2021,Simon1996}.
In addition, it has also been shown that, in the wolframite structure, the extent of the strain field associated with the substitution in the neighbouring octahedron in the zigzag chain is limited, because the coupling between the adjacent \ce{[$A$O6]} octahedral chains is weakened by the intervening \ce{[WO6]} octahedra \cite{Simon1996,Schofield1993}.
It should be noted, however, that -- apart from Cu--O pairs -- the average $A$--O distances in the MET and the HET are not very different from the $A$--O distances in pure \ce{$A$WO4} tungstates (cf. the positions of the vertical lines in Fig.\ \ref{fig5}).

\begin{figure}[t]
	\centering
	\includegraphics[width=0.95\textwidth]{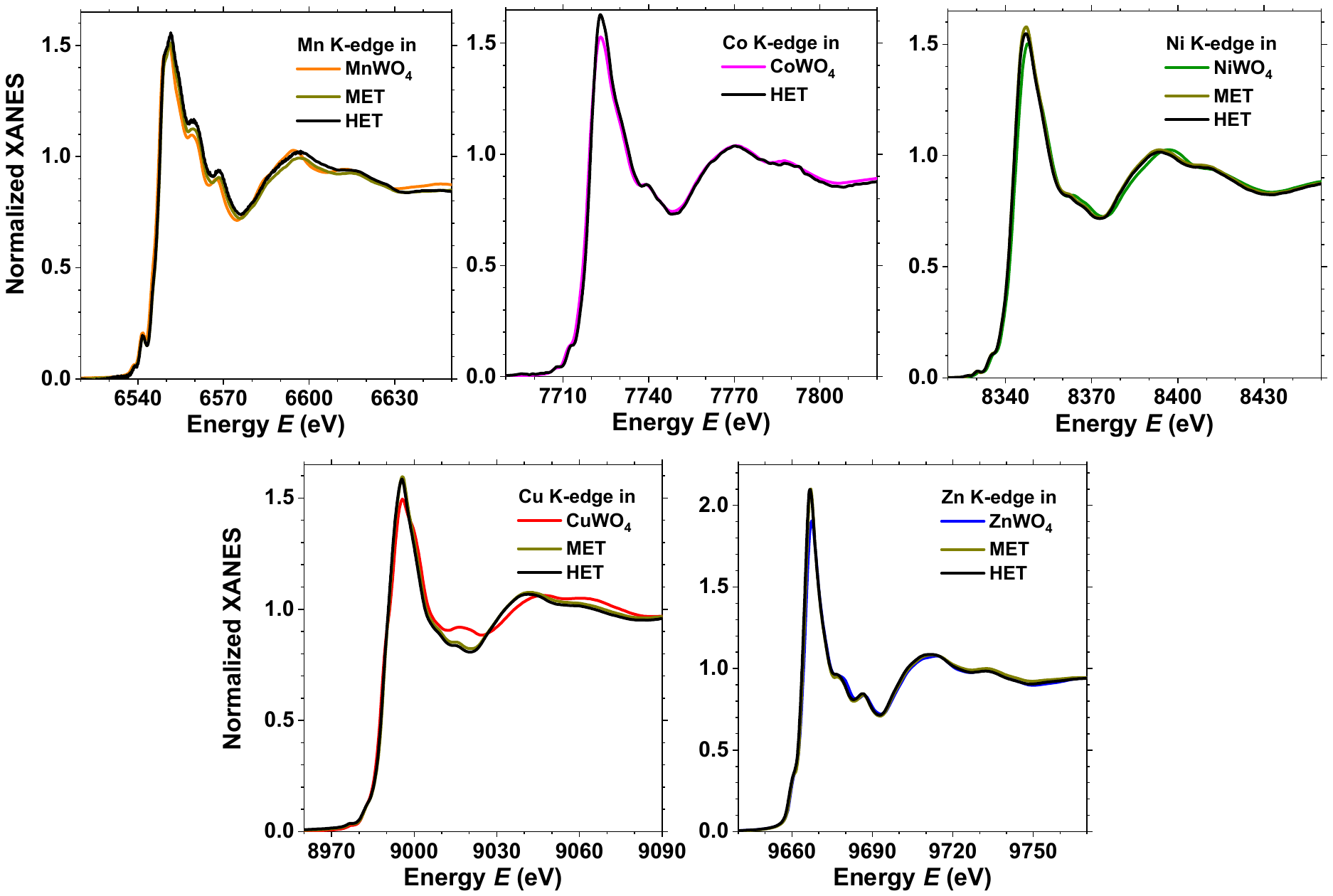}	
	\caption{X-ray absorption near-edge structure (XANES) spectra of Mn, Co, Ni, Cu, and Zn K-edges in pure tungstates \ce{$A$WO4},  medium-entropy tungstate (MET) \ce{(Mn,Ni,Cu,Zn)WO4}, and high-entropy tungstate (HET) \ce{(Mn,Co,Ni,Cu,Zn)WO4}. See the main text for details.}
	\label{fig6}
\end{figure}

Finally, we address the question of sample stoichiometry. As  noted in Refs. \cite{Bakradze2021,Schofield1993}, the stoichiometry of tungstates is difficult to demonstrate.
In \ce{$A$WO4} tungstates, $A$-type cations are in the +2 oxidation state, whereas W cations are in the +6 oxidation state; therefore, any non-stoichiometry would depend on the presence of multiple valency states of the $A$ and/or W cations.
Any non-stoichiometry or disorder in the tungstate samples would have to be a major phenomenon, but no signs of non-stoichiometry or disorder were found in XRD or XAS experiments.
The shape and position of the metal cation K-edge X-ray absorption near-edge structure (XANES) spectra in \ce{$A$WO4}, the MET, and the HET indicate a single type of $A$ octahedral site (Fig.\ \ref{fig6}).
The XANES spectra at the Mn, Co, Ni, and Zn K-edges show no pronounced differences in pure \ce{$A$WO4}, the MET, and the HET, thus implying similarity of the local atomic environment around each of the elements in these compounds.
Thus, unlike other HEMs, e.g. in \ce{(Co,Cu,Mg,Ni,Zn)_{1-c}Li_cO} \cite{Mozdzierz2021}, where the formation of oxygen vacancies was found in samples with Li content $c > 0.20$, we do not observe any prominent lattice distortion effect near 3d metal cations.
At the same time, the Cu K-edge XANES spectrum in \ce{CuWO4} differs from the spectra in the MET and the HET, evidencing different distortions of \ce{[CuO6]} octahedra because of the reduction of the JT axial distortion as shown by our RMC analysis (Fig.\ \ref{fig5}).

\section{Conclusions}\label{s:conc}
For the first time, an MET  and an HET -- \ce{(Mn,Ni,Cu,Zn)WO4} and \ce{(Mn,Co,Ni,Cu,Zn)WO4}, respectively -- were synthesized and characterized by powder XRD, Raman spectroscopy, and XAS.
The stoichiometric composition of the compounds was confirmed by chemical analysis. The XRD data confirmed the formation of a single-phase monoclinic ($P2$/$c$) material.
The room-temperature Raman scattering spectra demonstrated a strong composition dependence of the most intense $\mathrm{A_g}$ band in the series \ce{ZnWO4}, \ce{(Ni,Zn)WO4},  \ce{(Mn,Ni,Cu,Zn)WO4}, and \ce{(Mn,Co,Ni,Cu,Zn)WO4}.
The changes in vibrational dynamics were attributed to competing $A$--O and W--O interactions resulting in different W--O bond strengths in these materials.

Analysis of the low-temperature EXAFS spectra simultaneously at five  metal absorption edges in \ce{(Mn,Ni,Cu,Zn)WO4} and six metal absorption edges in \ce{(Mn,Co,Ni,Cu,Zn)WO4} using the RMC method allowed us to optimize atomic configurations on the basis of an ideal solid solution model and to study local static distortions around each metal cation.
The composition-induced distortions of \ce{[WO6]} and \ce{[$A$O6]} octahedra were evidenced by the shapes of W--O and $A$--O RDFs.
Similarly, as in pure wolframites \cite{Timoshenko2014,Bakradze2021}, the W--O RDF in \ce{(Mn,Ni,Cu,Zn)WO4} and \ce{(Mn,Co,Ni,Cu,Zn)WO4} exhibits two peaks caused by the second-order JT effect \cite{KUNZ1995}.
The formation of \ce{(Mn,Ni,Cu,Zn)WO4} and \ce{(Mn,Co,Ni,Cu,Zn)WO4} has a more pronounced effect on \ce{[$A$O6]} octahedra: the Co--O, Mn--O, and Zn--O RDFs are very broad; in contrast, the Ni--O RDF is very narrow.
This evidences the structural role of nickel cations and their strong tendency to organize the local environment, whereas other cations adapt to it.
An interesting result is related to the local structure of copper ions: a highly asymmetric shape of the Cu--O RDF differs from that expected for \ce{[Cu^{2+}O6]} octahedra in pure \ce{CuWO4}, i.e. in the case of a strong JT effect \cite{Timoshenko2014}.

To conclude, we have demonstrated that an accurate multi-edge analysis of EXAFS data by the RMC method  can reveal the peculiarities of local atomic environments in highly complex medium- and high-entropy, low-symmetry materials.

\section*{Declaration of competing interests}
The authors declare that they have no known competing financial interests or personal relationships that could have appeared to influence the work reported in this article.

\section*{Acknowledgements}
G. Bakradze acknowledges financial support provided by the Latvian Council of Science for project no. 1.1.1.2/VIAA/3/19/444 (agreement no. 1.1.1.2/16/I/001) realized at the Institute of Solid State Physics, University of Latvia. The Institute of Solid State Physics, University of Latvia, as a centre of excellence, has received funding from the European Union's Horizon 2020 Framework Programme H2020-WIDESPREAD-01-2016-2017-TeamingPhase2 under grant agreement no. 739508, project CAMART2.

\section*{Data availability}
Data will be made available on request.

\end{document}